\documentclass[twocolumn]{aastex631}

\hypersetup{
   colorlinks,
   linkcolor={blue!88!black!80},
   citecolor={blue!88!black!80},
   urlcolor={blue!88!black!80}}

\usepackage{color,subfigure,lineno}
\usepackage{float}
\usepackage{mathrsfs}
\usepackage{lineno}
\usepackage{mathrsfs}

\newcommand{\RNum}[1]{\uppercase\expandafter{\romannumeral #1\relax}}

\newcommand{\Hbeta}{H{$\beta$}}
\newcommand{\halpha}{H{$\alpha$}}

\def \OIII {[O\,{\sc iii}]}

\newcommand{\comments}[1]{}

\newcommand\ltday{\mathop{\mbox{light-day}}}

\begin{document}

\title{Estimating AGN Black Hole Masses via Continuum Reverberation Mapping in the Era of LSST}

\author[0000-0002-2052-6400]{Shu Wang}
\affiliation{Department of Physics \& Astronomy, Seoul National University, Seoul 08826, Republic of Korea; hengxiaoguo@gmail.com, wangshu100002@gmail.com}

\author[0000-0001-8416-7059]{Hengxiao Guo}
\affiliation{Key Laboratory for Research in Galaxies and Cosmology, Shanghai Astronomical Observatory, Chinese Academy of Sciences, 80 Nandan Road, Shanghai 200030, People's Republic of China}

\author[0000-0002-8055-5465]{Jong-Hak Woo}
\affiliation{Department of Physics \& Astronomy, Seoul National University, Seoul 08826, Republic of Korea}

\begin{abstract}
Spectroscopic reverberation mapping (RM) is a direct approach widely used to determine the mass of black holes (BHs) in active galactic nuclei (AGNs). However, it is very time consuming and difficult to apply to a large AGN sample. The empirical relation between the broad-line region size and luminosity (\Hbeta\ $R_{\rm BLR}-L_{\rm}$) provides a practical alternative yet is subject to large scatter and systematic bias. Based on a relation between the continuum emitting region (CER) size and luminosity ($R_{\rm CER}$--$L$) reported by Netzer (2022), we present a new BH mass estimator via continuum RM (CRM) by comparing $R_{\rm CER}$ and $R_{\rm BLR}$, assuming that the continuum lags are dominated by the diffuse continuum emission. Using a sample of 21 AGNs, we find a tight $R_{\rm BLR}$--$R_{\rm CER}$ relation (scatter$\sim$0.28 dex), and that $R_{\rm BLR}$ is larger than $R_{\rm CER}$ at 5100\AA\ by an average factor of 8.1. This tight relation enables the BH mass estimation based on the CRM combined with the velocity information. Applying the relation to rest objects in our CRM sample, we demonstrate that the predicted $R_{\rm BLR,CRM}$ follow the existing \Hbeta\ $R_{\rm BLR}-L_{\rm}$ relation well and the estimated CRM BH masses are consistent with the RM/SE BH masses using \Hbeta. This method will provide significant applications for BH mass estimation thanks to the short continuum lags and the easily accessible high-cadence, large-area photometric data, especially in the era of Legacy Survey of Space and Time.

\end{abstract}


\section{Introduction} \label{sec:intro}
Black hole (BH) mass is a key parameter in active galactic nucleus (AGN) studies, and is vital for understanding the coevolution of BHs and host galaxies \citep[e.g.,][]{Kormendy13}, the early formation history of supermassive BHs \citep[e.g.,][]{Inayoshi20}, and accretion physics \citep[e.g.,][]{Pringle81}. To accurately estimate the BH mass, one of the primary methods is observing the motion of stars or gas around the BHs to derive the dynamical mass, which is the most accurate approach for nearby BHs whose influence of sphere can be spatially resolved \citep[e.g.,][]{Ghez08,Hicks08,GRAVITYCollaboration19}. However, this approach is not accessible for more distant galaxies and is difficult to apply to a large sample. 

Reverberation mapping \citep[RM,][]{Blandford82} of AGN provides a unique chance to estimate the BH mass in distant galaxies. It measures the time delay ($\tau$) between the variability of the continuum and the broad-line emission to infer the broad-line region (BLR) size ($R_{\rm BLR}$= c$\tau$). Combined with the broad-line width ($\Delta V$) which is a proxy of the virial velocity of gas clouds in the BLR, we can estimate the BH mass following: 

\begin{equation}
M_{\rm BH} = \frac{f\Delta V^2 R_{\rm BLR}}{G}
\end{equation}
where $G$ is the gravitational constant and $f$ is the virial factor accounting for the unknown BLR orientation, kinematics, structure, and etc.. The traditional spectroscopic RM requires intensive observational resources and is usually very time-consuming, especially for luminous quasars at high redshift \citep[e.g.][]{Kaspi00, Grier17, Woo19b, Malik22}.

A scaling relation between the BLR size and the continuum luminosity ($R_{\rm BLR}$--$L$ relation) is established using \Hbeta\ based on $\sim40$ local AGNs \citep[e.g.,][]{Kaspi00,Bentz13}. Its best-fit slope is $0.533^{+0.035}_{-0.033}$ \citep{Bentz13}, fully consistent with the expectation from simple photoionization models. According to the definition of the ionization parameter $U_{\mathrm{ H}} = \frac{Q(H)}{4\pi R^{2}n_{\mathrm{ H}}c}$ \citep{Davidson72b}, 
where $Q(H)$ is the  production rate of hydrogen-ionizing photons that is proportional to the $L$, and $n(H)$ is the hydrogen density of the gas, the size $R$ that is responsible for the line emission will be proportional to $\sqrt{U_{\mathrm{ H}}n_{\mathrm{ H}} L}$. The observed \Hbeta\ $R_{\rm BLR}$--$L$ relation is a natural consequence if $U_{\mathrm{ H}}$ and $n_{\mathrm{ H}}$ of \Hbeta-emitting regions are similar among different AGNs.

The $R_{\rm BLR}$--$L$ relation enables the BH mass estimation using single-epoch (SE) luminosity and broad-line width, which can be easily applied to a large sample of AGNs \citep[e.g.,][]{Vestergaard06,Shen11,Liu19,Rakshit20,Wu22}. However, the current $R_{\rm BLR}$--$L$ relation becomes much more complex and less tight than the canonical one \citep{Bentz13}. AGNs with super-Eddington accretion rates have significantly smaller BLR size \citep[e.g.,][]{Du15,Du16,Du18b} than the expectation from canonical $R_{\rm BLR}$--$L$ relation, which results in overestimation of their SE BH mass using traditional estimators \citep{Du19}. 

Apart from the BLR, the more extended dusty torus is also found to obey a similar $R$--$L$ relation \citep{Suganuma06,Koshida14,Lyu19}. Recently, \citet{GRAVITYCollaboration23} confirmed that the size of hot dust continuum $R_{\rm dust}$ measured using optical/near-infrared interferometry is tightly correlated with the $R_{\rm BLR}$, with an intrinsic scatter of 0.25 dex. This suggests that the BH mass can be estimated via the hot dust continuum size combined with a broad-line width, although the slope of $R_{\rm dust}-L$ relation is 
 found slightly flatter than 0.5. This method is currently limited to the brightest objects with $K$ band brighter than 11 mag. A large sample study will be feasible using the next-generation instrument with upgraded sensitivity and sky coverage \citep{GRAVITY+Collaboration22}. 

The optical continuum RM (CRM) programs have also made significant progress in resolving the continuum emitting region (CER) sizes of a number of local AGNs with observations through X-ray to UV/optical.
\citep[e.g.,][]{Fausnaugh16,Edelson17,Cackett18,Cackett20,HernandezSantisteban20,Vincentelli21,Kara21}. Several other works successfully measured the optical continuum lags using modern photometric time-domain surveys \citep{Jiang17,Mudd18,Homayouni19,Yu20b,Jha22,Guo22a, Guo22b}. A key result from these CRM campaigns is that the observed CER sizes ($R_{\rm CER}$) are about 3 times larger than the prediction of the standard thin disk model \citep[SSD,][]{Shakura73}. One of the popular explanations suggests that the diffuse continuum (DC) emission from the BLR dominates the observed continuum lags \citep[e.g.,][]{Korista01,Korista19,Lawther18,Netzer22}. This idea is strongly supported by the $U/u$ band lag excess in  the lag spectrum as a function of wavelength \citep[e.g., NGC 4593,][]{Cackett18}, although not every object present this feature \citep[e.g., Mrk~817,][]{Kara21}.

Based on several local AGNs, \citet{Netzer22} proposed the $R_{\rm CER}$--$L$ relation and suggested that the $R_{\rm CER}$--$L$ relation is likely a down-scale version of the \Hbeta\ $R_{\rm BLR}$--$L$ relation. This result is quickly confirmed by \citet[][hereafter G22]{Guo22a} with a much larger sample of AGNs using the light curves from Zwicky Transient Facility (ZTF) and is also extended to the low mass regime \citep{Montano22}. 

In this Letter, we present the tight scaling relation between $R_{\rm CER}$ and $R_{\rm BLR}$ and propose a new method to estimate the BH mass via CRM. We describe the sample in \S \ref{sec:sample} and investigate the feasibility of CRM BH mass estimation in \S \ref{sec:results}. We discuss the advantages and limitations of this new approach, as well as future prospects in the era of Legacy Survey of Space and Time (LSST) in \S \ref{sec:diss}. Finally, we draw our conclusions in \S \ref{sec:con}. Throughout this paper, we use the $\Lambda$CDM cosmology, with $H_{\rm 0}$ = 72.0, and $\Omega_{\rm m}$ = 0.3.

\section{Data and Sample} \label{sec:sample}

\begin{table*}[htbp]
    \centering
    \caption{Sample properties}    \label{tab:info}
    \begin{tabular}{@{\extracolsep{6pt}}l c c c c c l c c}
    \hline \hline
     Object name  &  $z$ & $R_{\rm CER}$ & Sample & Ref.$\,1$ & ${R_{\rm BLR}}^a$ & log$L_{\rm 5100}$ & log$\lambda_{\rm Edd}$ & Ref.$\,2$  \\ 
  
    &   & (light-day) &  &  & (light-day) & (erg s$^{-1}$) & &  \\ 
     
     \hline

Ark 120 & 0.033 & 2.48$\pm$0.57 & Local & 1 & 39.5$^{+8.5}_{-7.8}$ & 43.81$\pm$0.25 & $-1.90$ & 13 \\ 
Fairall 9 & 0.047 & 4.35$\pm$0.23 & Local & 2 & 17.4$^{+3.2}_{-4.3}$ & 43.92$\pm$0.05 & $-1.43$ & 14,15 \\ 
MCG +08-11-011 & 0.021 & 1.14$\pm$0.10 & Local & 3 & 15.7$^{+0.5}_{-0.5}$ & 43.28$\pm$0.05 & $-1.70$ & 16,15 \\ 
Mrk 110 & 0.035 & 0.88$\pm$0.10 & Local & 4 & 25.6$^{+8.9}_{-7.2}$ & 43.60$\pm$0.02 & $-0.78$ & 13 \\ 
Mrk 142 & 0.045 & 1.16$\pm$0.05 & Local & 5 & 6.4$^{+7.3}_{-3.4}$ & 43.53$\pm$0.04 & $-0.16$ & 13 \\ 
Mrk 509 & 0.026 & 2.95$\pm$0.36 & Local & 6 & 79.6$^{+6.1}_{-5.4}$ & 44.13$\pm$0.03 & $-1.28$ & 17,15 \\ 
Mrk 817 & 0.031 & 3.26$\pm$0.28 & Local & 7 & 19.9$^{+9.9}_{-6.7}$ & 43.68$\pm$0.09 & $-1.62$ & 13,15 \\ 
NGC 2617 & 0.014 & 0.54$\pm$0.10 & Local & 3 & 4.32$^{+1.10}_{-1.35}$ & 42.61$\pm$0.10 & $-2.38$ & 16,15 \\ 
NGC 4151 & 0.003 & 1.27$\pm$0.41 & Local & 8 & 6.82$^{+0.48}_{-0.57}$ & 42.31$\pm$0.06 & $-2.50$ & 16,15 \\ 
NGC 4395 & 0.001 & 0.0130$\pm$0.0005 & Local & 9 & $0.058^{+0.010}_{-0.010}$  & 39.76$\pm$0.06  & $-1.56$ & 18,19 \\
NGC 4593 & 0.009 & 0.52$\pm$0.08 & Local & 10 & 4.0$^{+0.8}_{-0.7}$ & 42.56$\pm$0.37  & $-2.01$ & 13 \\ 
NGC 5548 & 0.017 & 2.12$\pm$0.03 & Local & 11 & 13.9$^{+11.2}_{-6.2}$ & 43.24$\pm$0.19 & $-2.17$ & 13 \\ 
SDSS J121752.16$+$333447.2 & 0.178 & 4.06$\pm$1.35 & Parent & 12 & 26.5$^{+21.2}_{-20.7}$ & 44.20$\pm$0.02 & $-1.09$ & 20 \\ 
SDSS J152624.02$+$275452.1 & 0.231 & 7.35$\pm$0.21 & Parent & 12 & 63.9$^{+10.3}_{-9.3}$ & 44.82$\pm$0.01 & $-0.85$ & 20 \\ 
PG 0049+171 & 0.064 & 7.04$\pm$1.03 & Core & 12 & 39.5$^{+3.3}_{-2.6}$ & 44.04$\pm$0.06$^b$ & $-1.36$ & 21 \\ 
PG 1048+342 & 0.167 & 12.13$\pm$2.20 & Core & 12 & 36.8$^{+2.4}_{-3.4}$ & 44.49$\pm$0.05$^b$ & $-0.55$ & 21 \\ 
PG 1402+261 & 0.164 & 4.32$\pm$1.79 & Parent & 12 & 95.9$^{+7.1}_{-23.9}$ & 44.74$\pm$0.05 & $-0.42$ & 22 \\ 
PG 1426+015 & 0.086 & 15.52$\pm$3.44 & Parent & 12 & 95.0$^{+29.9}_{-37.1}$ & 44.57$\pm$0.02 & $-1.66$ & 23,15 \\ 
PG 1501+106 & 0.036 & 8.25$\pm$0.47 & Core & 12 & 26.0$^{+2.0}_{-2.0}$ & 43.95$\pm$0.06$^b$ & $-1.63$ & 21 \\ 
PG 1552+085 & 0.119 & 6.13$\pm$1.40 & Parent & 12 & 25.0$^{+12.6}_{-12.0}$ & 44.29$\pm$0.05 & $-0.24$ & 22 \\ 
PG 1626+554 & 0.132 & 4.19$\pm$1.84 & Core & 12 & 77.1$^{+5.5}_{-2.6}$ & 44.61$\pm$0.06  & $-1.10$ & 22 \\

\hline
\multicolumn{9}{p{0.99\textwidth}}{{\bf Notes.} 

a. The $R_{\rm BLR}$ are represented by \Hbeta\ except for NGC~4395 of which only \halpha\ $R_{\rm BLR}$ is available.

b. The log$L_{5100}$ are the total luminosity at 5100 \AA\ including the host contamination. 

{\bf References 1} (reference of CER size $R_{\rm CER}$ characterized at rest-frame 5100\AA): (1) \citet{Lobban20}; (2) \citet{HernandezSantisteban20} (3) \citet{Fausnaugh18} (4) \citet{Vincentelli21} (5) \citet{Cackett20} (6) \citet{Edelson19} (7)  \citet{Kara21} (8) \citet{Edelson17} (9) \citet{Montano22} (10) \citet{Cackett18} (11) \citet{Fausnaugh16}. (12) \citet{Guo22a}. {\bf References 2} (reference of BLR size $R_{\rm BLR}$ and AGN luminosity log$L_{5100}$): (13) \citet{Du19}  (14) \citet{Santos-Lleo97} (15) \citet{DallaBonta20} (16)\citet{Fausnaugh17} (17) \citet{Peterson98a}  (18) \citet{Woo19} (19) \citet{Cho20}. (20) Woo et al. (2023, in preparation) (21) \citet{Bao22} (22) \citet{Hu21} (23) \citet{Kaspi00} . For objects with multiple BLR RM measurements, we use the averaged $R_{\rm BLR,H\beta}$ and  log$L_{5100}$ from the Table 1 by \citet{Du19}.

}

\end{tabular}
\end{table*}

The CRM sample is mainly from G22 who performed a uniform lag analysis for a large sample of AGNs using \emph{gri}-band light curves from ZTF \citep{Bellm19}. ZTF is a time-domain survey that utilizes a 1.2 m telescope at the Palomar Observatory to scan the entire visible sky (DEC $\geq-30^{\circ}$)  with a cadence of $\sim3$ days since 2018. G22 cross-matched the Million Quasar Catalog \citep{Flesch21} with ZTF Data Release 7 that contains the PSF-based photometric light curves from March 2018 to June 2021. A total of 455 spectroscopically confirmed type-\RNum{1} AGNs were selected at redshift $z<0.8$ to have well-sampled light curves ($N_{\rm epoch}>20$ for each of the three bands) and show good cross-correlation between $g$ and $r$ band light curves (see section 2.1.1 in G22 for details).

The lags of these 455 AGNs were measured using both the interpolated cross-correlation function \citep[ICCF,][]{Peterson98b,Sun18} and {\tt JAVELIN} \citep{Zu11}. The $R_{\rm CER}$ were estimated by fitting a power-law function $R_{\rm CER} =R_{\rm CER,0}\,\lambda^{\beta}$ to the inter-band ($g$-$r$, $g$-$i$) lag-wavelength relation. The power-law slope $\beta$ was fixed to $\beta=4/3$ as suggested by previous AGN CRM studies \citep{Fausnaugh16,Jiang17,Yu20b,Homayouni19,Homayouni22}. The $R_{\rm CER}$ at rest-frame 2500~\AA\ are available in the Table 4 of G22. 

We use the high-quality sub-samples selected by G22 based on lag uncertainties, consistency between two lag measuring approaches, lag reliability, etc. A total of 94 objects with reasonably good lag quality were selected as the parent sample, from which 38 objects with the most confident lag measurements were labeled as the core sample. Note that in this work, we use the parent sample to refer to the 56 objects excluding the core sample objects ($94-38=56$) and the core sample still represents the best 38 objects. We adopt $R_{\rm CER}$ based on ICCF measurements. Using {\tt JAVELIN} results will not change our conclusion since the lag difference between ICCF and {\tt JAVELIN} is required to be small for these two samples.

We supplement G22 CRM sample with 12 local AGNs \citep{Fausnaugh16,Fausnaugh18,Edelson17,Cackett18,Cackett20,HernandezSantisteban20,Vincentelli21,Kara21,Montano22}, which is denoted as the local sample. Most of these AGNs are extensively monitored at multi-wavelengths ranging from X-ray to near-infrared, providing the best constraints on the $R_{\rm CER}$ to date. Their $R_{\rm CER}$ at rest-frame 2500 \AA\ are summarized in the Table 3 of G22. Finally, our CRM sample consists of 94$+12=$106 objects.

We cross-match the CRM sample with existing \Hbeta\ RM databases \citep{Bentz15,Du19,DallaBonta20} and new RM measurements after these works \citep[e.g.,][Woo et al., in preparation]{Hu21,Cho21,U22,Bao22}. In total there are 21 objects successfully cross-matched, among which there are 4, 5, and 12 objects from the core, parent sample, and local sample, respectively. For each object, we collect the \Hbeta\ BLR size, optical luminosity $\lambda L_{\lambda}$(5100\AA). Note that if there are multiple measurements of individual object, we use the average BLR size and luminosity provided by \citet{Du19}. The luminosity is transferred into the same cosmology if necessary. Table \ref{tab:info} lists the properties for all the cross-matched objects.

\section{Results}\label{sec:results}

\begin{figure*}[h]
\centering
\includegraphics[width=0.9\textwidth]{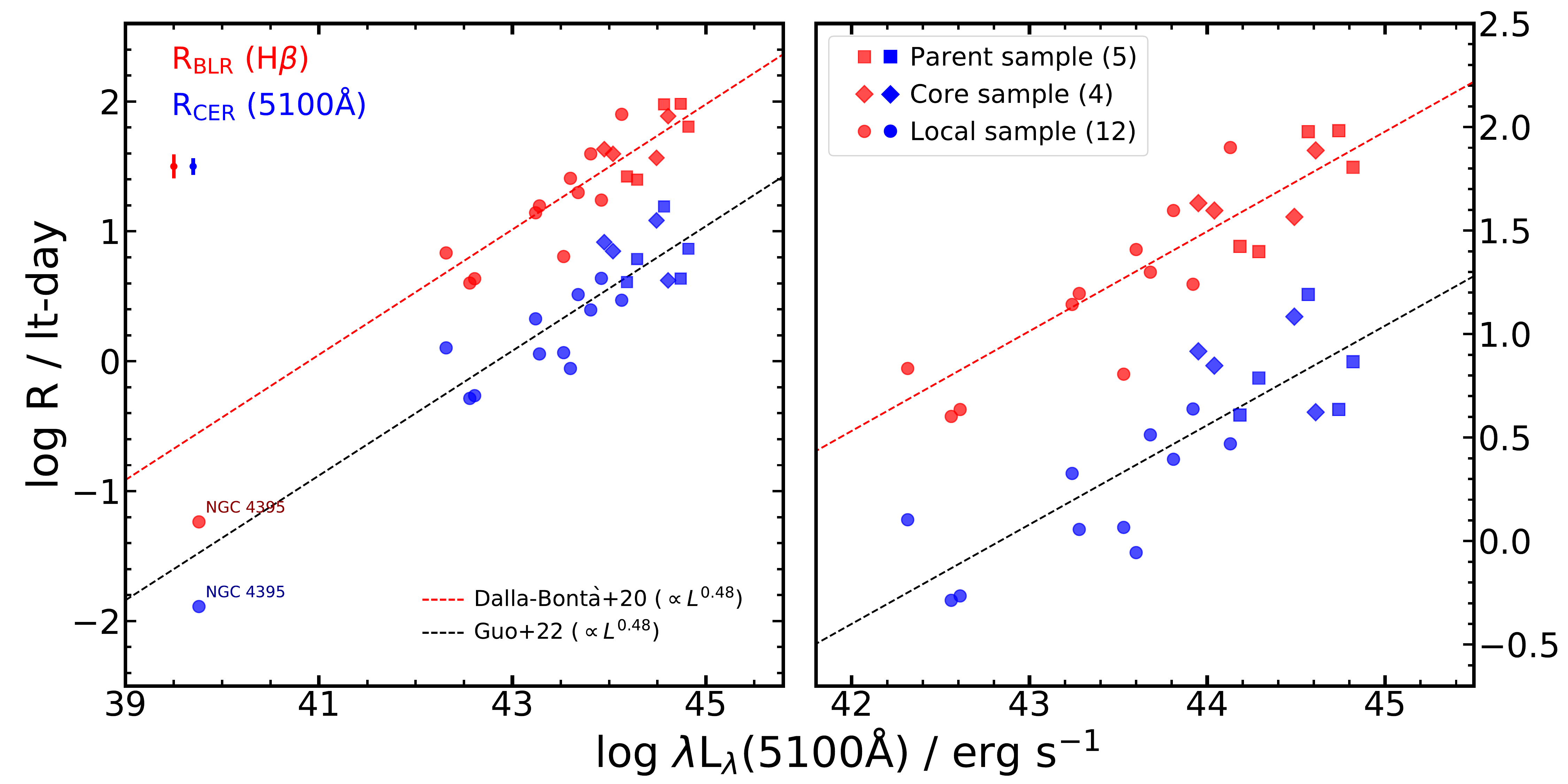}
\includegraphics[width=0.9\textwidth]{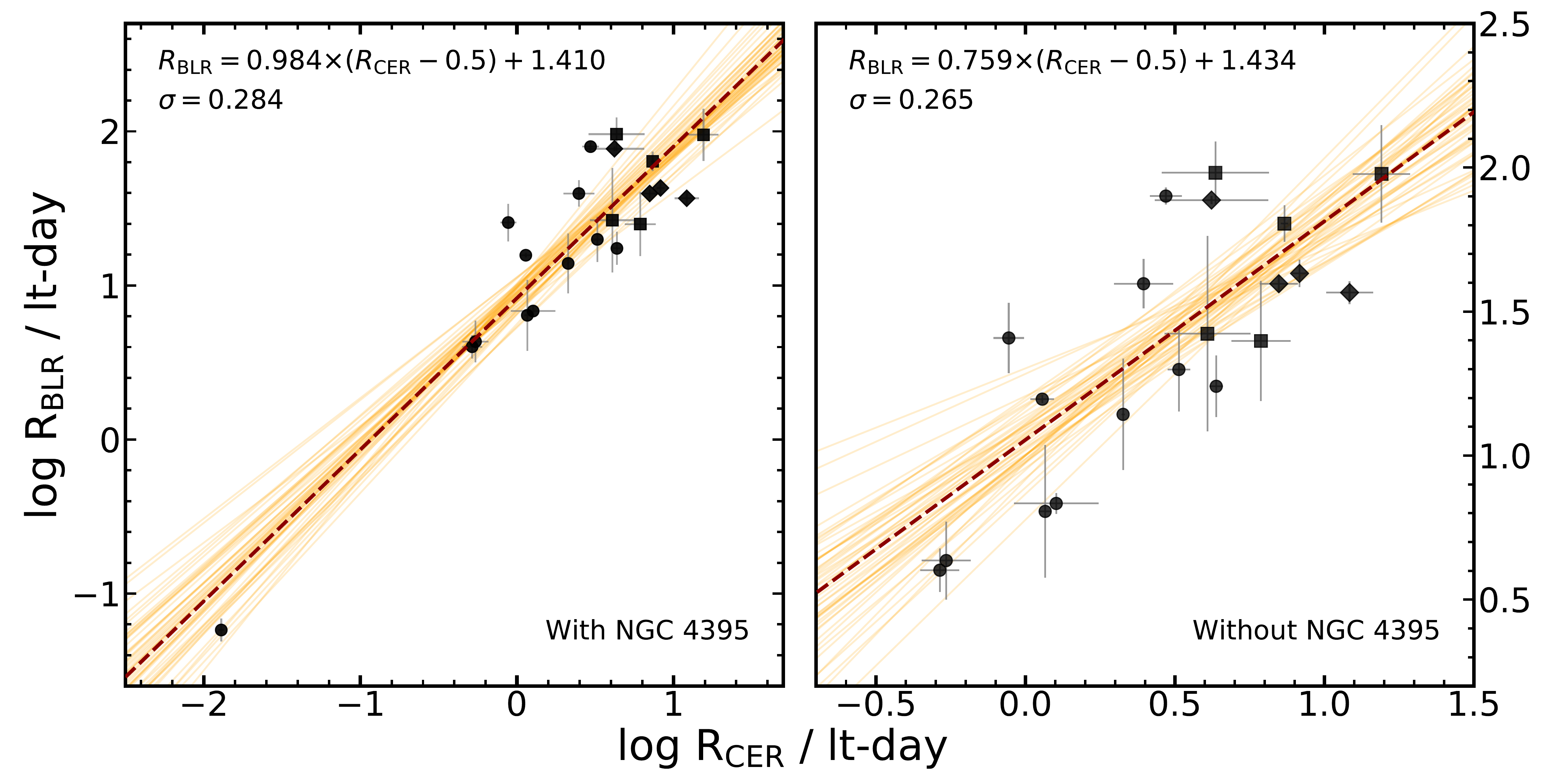}

\caption{Upper panels: comparison between the BLR and CER size--luminosity relation. The red symbols represent the \Hbeta\ BLR sizes ($R_{\rm BLR}$) except for NGC~4395 where only \halpha\ is available, while the blue symbols represent the CER sizes ($R_{\rm CER}$) at rest-frame $\lambda=5100$\AA. Parent, core, and local sample are displayed by squares, diamonds, and circles. The CER size has a luminosity dependence of $R_{\rm CER}\propto L^{0.48}$ (G22), comparable to that of BLR, i.e., $R_{\rm BLR}\propto L^{0.48}$ \citep{DallaBonta20}. Lower panels: the BLR--CER sizes comparison using 21 objects with both BLR and CER size well measured. The symbols are same as two upper panels. The brown dashed line represents the best-fit relation from linear regression and the orange lines are randomly selected realizations in the MCMC chains derived by {\tt emcee}. The left two panels show the results including NGC~4395 while the two right panels show the results without NGC~4395. }
\label{fig:R-L}
\end{figure*}

The upper panels in Figure \ref{fig:R-L} compare the BLR and CER $R$--$L$ relation. The BLR sizes are characterized by the most well-studied emission line \Hbeta, except for NGC~4395 where only \halpha\ is available. In this section, we will investigate the connection including and excluding NGC~4395, respectively. The $R_{\rm CER}$ is characterized at rest-frame 5100$\,$\AA\ which is transferred from 2500$\,$\AA\ assuming the lag-wavelength power-law index $\beta=4/3$ \citep[i.e., $R_{\rm CER,5100}=R_{\rm CER,2500}\left(5100/2500\right)^{\frac{4}{3}}$,][G22]{Fausnaugh16,Homayouni22}. 

A remarkable feature of Figure \ref{fig:R-L} is that the two $R$--$L$ relations are approximately parallel. The \Hbeta\ $R_{\rm BLR}$--$L$ relation has a slope of $0.533^{+0.035}_{-0.033}$, derived by \citet{Bentz13} based on a sample of spectroscopically monitored AGNs with luminosity well constrained using HST images. Recently, \citet{DallaBonta20} updated the slope to $0.482\pm0.029$ based on the extended database. On the other hand, the $R_{\rm CER}$--$L$ relation has a slope of $0.48\pm0.04$, established using 49 objects (see G22, 38 objects from the core sample, and 12 objects from the local sample). The two slopes are fully consistent with each other but the CER size is substantially smaller. Another interesting fact is that both the $R_{\rm BLR}$ and $R_{\rm CER}$ of NGC~4395 show substantial offset to the best-fit relation of more luminous AGNs. 

The similar dependency of $R_{\rm CER}$ and $R_{\rm BLR}$ on luminosity can be interpreted as a significant contribution from the BLR DC in the optical continuum band \citep[e.g.,][G22]{Li21,Netzer22}. If this scenario is the real case, this similarity provides an opportunity to use the CRM derived $R_{\rm CER}$ as a surrogate of the spectroscopic RM measured $R_{\rm BLR}$ to estimate the BH mass. To study this feasibility, we plot the direct comparison between $R_{\rm BLR}$ and $R_{\rm CER}$ in the lower panels in Figure \ref{fig:R-L} and fit the relation using the following equation:

\begin{equation} 
    {\rm log}\, (R_{\rm BLR}/{\ltday}) = \alpha \,{\rm log}\, (R_{\rm CER}/R_{\rm CER,0}) + K \label{equ:R-R}
\end{equation} 
where $\alpha$ and $K$ are the slope and intercept, respectively. $R_{\rm CER,0}$ is the reference point, which is set to  log$R_{\rm CER,0}=0.5$, close to the median of our cross-matched sample. We perform linear regression using {\tt emcee} \citep{Foreman-Mackey13} with the likelihood expressed by:
\begin{equation}
\ln \mathscr{L}^2=-\frac{1}{2}\sum_{i=1}^{N} \left( \frac{(y_i-m_i)^2}{s_{i}^{2}} + \ln(2\pi s_{i}^{2}) \right)
\end{equation}
where $y_i$ is the $i$th observed log$R_{\rm BLR}$, $m_i$ is the model prediction based on $i$th log$R_{\rm CER}$ through equation \ref{equ:R-R}, and $s_{i}^2=(\alpha x_{{\rm err},i})^2+y_{{\rm err},i}^2 + \sigma_{\rm int}^2$ involves the uncertainties of $R_{\rm CER}$, $R_{\rm BLR}$ and the intrinsic scatter $\sigma_{\rm int}$. We perform the fitting with and without NGC~4395, respectively. The best-fit slopes and intercepts as well as their uncertainties are summarized in Table \ref{tab:fitting_para_R-R_relation}. 

As shown by the lower panels of Figure \ref{fig:R-L}, the correlation between $R_{\rm BLR}$ and $R_{\rm CER}$ has a slope of $\alpha=0.984^{+0.101}_{-0.101}$ if NGC~4395 is included. Otherwise, it shows a slightly flatter slope of $\alpha=0.759^{+0.150}_{-0.157}$, which is still consistent with linear relation in normal space ($\alpha=1.0$) within 2$\sigma$ uncertainty. We adopt the fitting result with NGC~4395 as the fiducial relation because he inclusion of NGC~4395 provides much larger dynamical range thus better constraints of the slope.

The best-fit intercept of the fiducial case is $K=1.410^{+0.501}_{-0.348}$ at $R_{\rm CER,0}=0.5$, which reveals that the $R_{\rm BLR}$ is 8.1 times ($0.910$ dex) larger than the $R_{\rm CER}$. In the case of excluding NGC~4395, the result is 8.6 times ($0.934$ dex). These results are consistent with the predicted value from the radiation pressure confined (RPC) cloud model \citep{Netzer22} where the $R_{\rm BLR}/R_{\rm CER}\sim8.7$ (0.939 dex) assuming that continuum lags are fully contributed by the DC component and the BLR covers 20\% percent of the sky of the central disk. However, the broad-line RM usually assumes that the continuum is emitted from a very compact region, whose size is considered negligible when compared to $R_{\rm BLR}$. However, if the $R_{\rm CER}$ is actually $\sim$1/8 of the size of \Hbeta\ emitting region, it implies that the previous RM may have underestimated the BH mass by an average of 12.5\% using 5100\AA\ continuum. This underestimation can be even larger if the continuum is close to the Balmer limit,  e.g., using $B$ or $g$ band as the reference continuum for objects at redshift $z\sim0.2$ to $z\sim0.6$.

\begin{table}[htbp]
    \centering
    \caption{Fitting results of $R_{\rm BLR}$--$R_{\rm CER}$ relation}
    \label{tab:fitting_para_R-R_relation}
    \small
    \begin{tabular}{@{\extracolsep{0pt}}l c c c }
    \hline \hline 
    Sample &    $\alpha$  & $K$  & $\sigma$  \\ \hline
    With NGC~4395 &  $0.984^{+0.101}_{-0.101}$ & $1.410^{+0.501}_{-0.348}$ & $0.284^{+0.058}_{-0.054}$ \\
    Without NGC~4395  &  $0.759^{+0.150}_{-0.157}$ & $1.434^{+0.744}_{-0.604}$ & $0.265^{+0.056}_{-0.052}$ \\

    \hline
    \multicolumn{4}{p{0.46\textwidth}}{{\bf Notes.} These two fitting results represent the relation shown in the lower left and right panel of Figure \ref{fig:R-L}, respectively. We adopt the fitting result with NGC~4395 as the fiducial relation.}
    
    \end{tabular}

\end{table}

\begin{figure*}[htbp]
\centering
\includegraphics[width=0.49\textwidth]{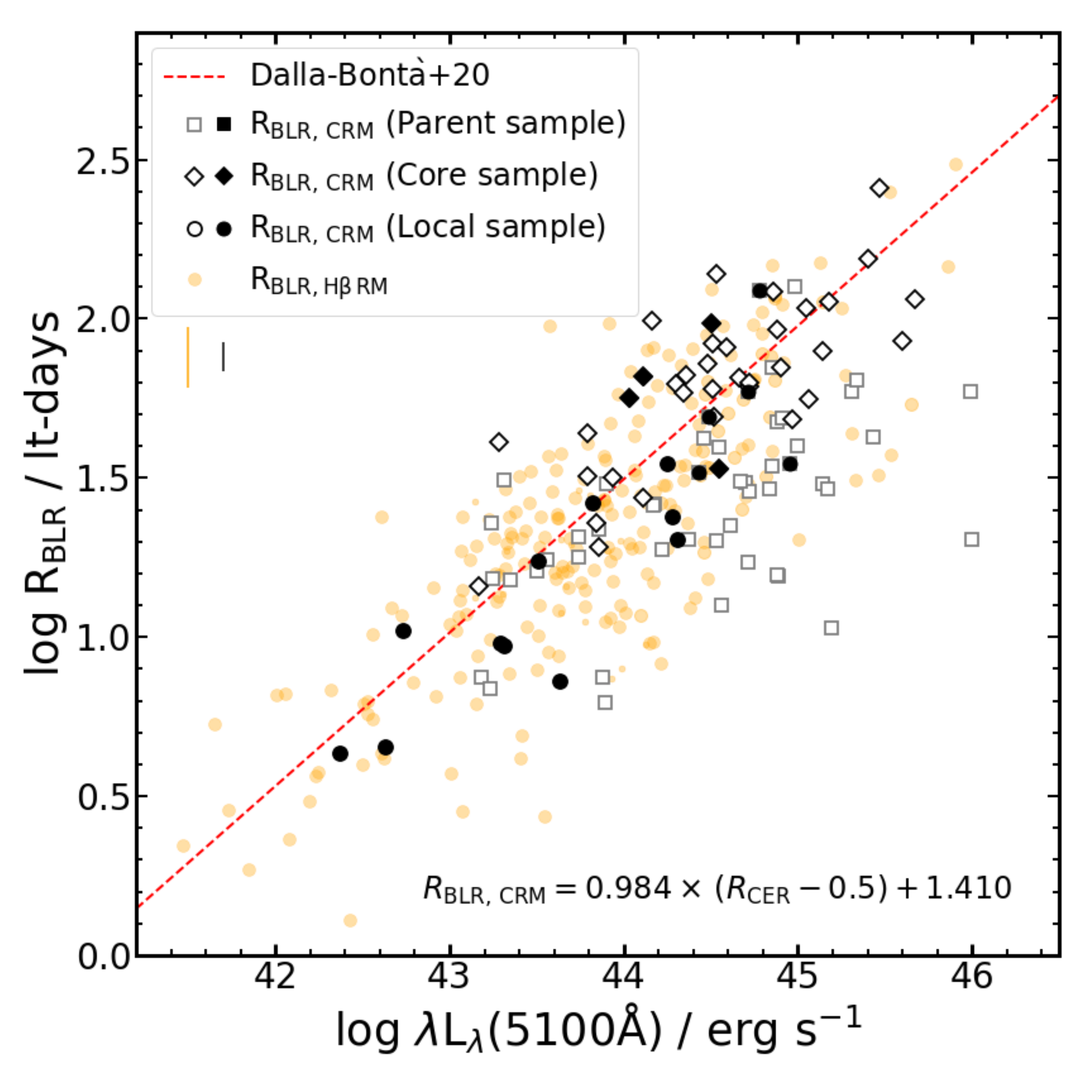}
\includegraphics[width=0.49\textwidth]{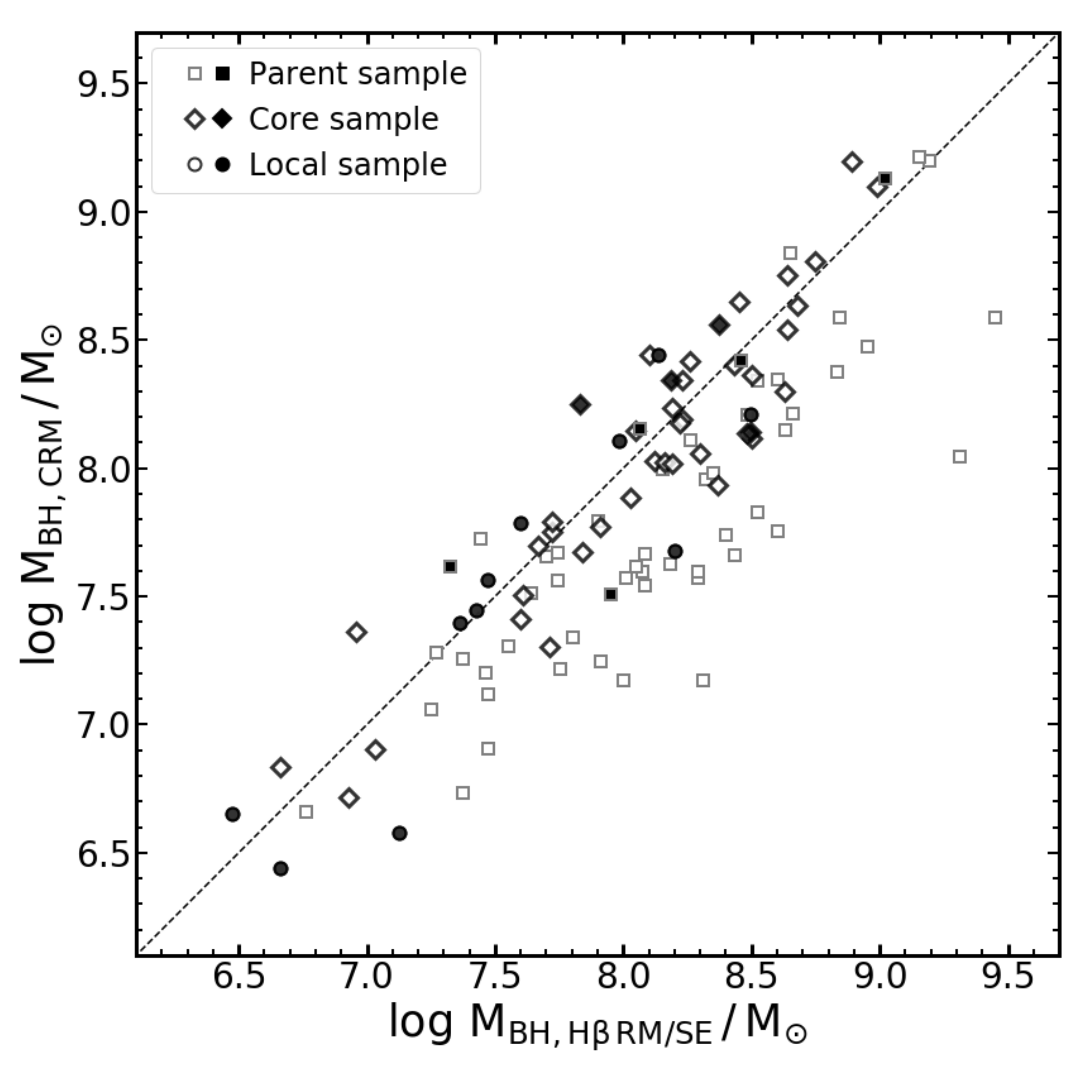}

\caption{Left panel: Application of the fiducial $R_{\rm BLR}$--$R_{\rm CER}$ relation to the rest objects in our CRM sample to predict the $R_{\rm BLR}$. Different samples are labeled using different symbols as in Figure \ref{fig:R-L}. Filled and unfilled black symbols represent the objects with and without available BLR size measurements, respectively. The light orange circles refer to the $R_{\rm BLR}$ of \Hbeta\ from spectroscopic RM observation. The red dashed line shows the best-fit relation of \Hbeta\ $R_{\rm BLR}$--$L$ relation by \citep{DallaBonta20}. Right panel: Application of the fiducial $R_{\rm BLR}$--$R_{\rm CER}$ relation to the rest objects in our CRM sample to calculate the CRM BH mass, compared with their \Hbeta\ RM (filled symbols) / SE (unfilled symbols) BH mass. For cross matched sample, both their RM and SE mass are included in comparison. The black dashed line indicates the 1:1 correspondence. The CRM and \Hbeta\ RM/SE BH mass show good consistency.  }\label{fig:appliation-R-L}
\end{figure*}

The intrinsic scatter of the fiducial case is $\sigma_{\rm int}=0.284$ dex, comparable to the current scatter of $R_{\rm BLR}$--$L$ relation \citep[$\sim$0.22 dex][]{DallaBonta20,Malik22}. In the case of excluding NGC~4395, the intrinsic scatter is even lower ($\sigma_{\rm int}=0.265$~dex). This scatter can be partly resulted from different contributions of BLR DC relative to the disk component among different objects if we assume DC is the main reason leading to this relation. The AGN variability between the BLR RM and CRM campaigns can also introduce a large scatter. In our sample, all $R_{\rm CER}$ were measured within recent $\sim7$ years, while the $R_{\rm BLR}$ of 7 out of 21 objects include measurements more than 20 years ago \citep[e.g.][]{Peterson98a, Kaspi00}. However, a direct comparison between the luminosity used in CRM studies and that reported by \Hbeta\ RM campaign shows no significant offset, indicating no dramatic variability for objects in our sample. Other factors can also contribute to the scatter. For example, G22 measures the $R_{\rm CER}$ using three optical bands but the local sample uses more bands, especially in the UV wavelengths. A simple combination can lead to extra scatter. 

Based on the $R_{\rm BLR}$--$R_{\rm CER}$ correlation, we can estimate the $R_{\rm BLR}$ from CRM, i.e., $R_{\rm BLR, CRM}$. Using the fiducial relation, we predict the $R_{\rm BLR}$ for the rest objects in the core and parent sample without  BLR size measurements. As shown in Figure \ref{fig:appliation-R-L}, the objects in the core sample (filled and empty diamonds) perfectly follow the trends of the existing \Hbeta\ $R_{\rm BLR}$--$L$ relation, exhibiting a similar scatter as direct \Hbeta\ RM measurements \citep[0.22 dex,][]{DallaBonta20,Malik22}. On the other hand, some objects in the parent sample (filled and empty squares) show smaller BLR size than the prediction from the canonical $R_{\rm BLR}$--$L$ relation \citep{DallaBonta20}, which is inherited from G22. It can be explained by the fact that the parent sample in G22 includes some small lag measurements that are consistent with zero within 1$\sigma$ (less reliable). In addition, the generally lower fractional variability (e.g., weaker variability with relatively larger photometric noise) of the parent sample can lead to an underestimation of the lag as found in the two-night observation of NGC~4395 \citep{Montano22}.

Next, we estimate the CRM BH mass combining the $R_{\rm BLR, CRM}$ and the FWHM from SE spectra in the literature \citep[e.g.,][]{Shangguan18,Du19,Liu19}. We compare the CRM BH mass with RM/SE mass using \Hbeta\ in the right panel of Figure \ref{fig:appliation-R-L} and find that they are in good agreement with each other. The median and $1\sigma$ scatter of the ratio log$\,({\rm M}_{\rm BH,CRM}/$M$_{\rm BH,SE})$ is $-0.03$ and 0.23 dex for core$+$local sample, respectively, confirming the feasibility of CRM BH mass estimation. Compared with the $\sim$0.28 dex intrinsic scatter of $R_{\rm BLR}$--$R_{\rm CER}$ relation, the $\sim$0.4 dex intrinsic scatter of virial factor $f$ \citep{Woo10,Woo13,Ho14} is still the primary source of the CRM BH mass uncertainty, which includes the intrinsic scatter of $M_{\rm BH}$--$\sigma_{*}$ relation \citep{Ferrarese00,Gebhardt00} as well as the diversity of BLR orientation and dynamic structures. The $f$ factor varies between different bulge types \citep{Ho14} and is suggested to correlate with BLR orientation \citep{Mejia-Restrepo18, Yu19}. The use of SE line width can introduce additional uncertainties due to asynchronous measurements of line width and the $R_{\rm CER}$ as well as the difference between SE/mean spectrum and rms spectrum \citep{Peterson04,Collin06,Wang19,Yu19}. Without determining $f$ factor for individual object as done in dynamical modelling using spectroscopic RM data \citep{Pancoast14,Williams18,Villafana22}, the CRM BH mass uncertainty is not likely to be lower than 0.4 dex.

Next, we explore the BH mass of NGC~4395. If adopting the fiducial relation, the estimated $R_{\rm BLR}$ of NGC~4395 from CRM is $166\pm6$ minutes with a $0.284$ dex intrinsic scatter, and the CRM BH mass is $(3.5\pm0.4)\times 10^{4}M_{\odot}$ with a $\sim$0.5 dex intrinsic scatter. This BH mass is more consistent with the RM BH mass \citep{Woo19,Cho21} relative to the dynamical BH mass \citep{Denbrok15}. 

Finally, we investigate the affection of accretion properties on the relation, calculating the Eddington ratios ($\lambda_{\rm Edd}$) as well as the dimensionless accretion rates \citep[$\dot{\mathscr{M}}$,][]{Wang14,Du15} of our sample. None objects in our sample exhibit an Eddington ratio $\lambda_{\rm Edd}>1$ (see Table \ref{tab:info}) while if using  $\dot{\mathscr{M}}\sim3$ as the criterion to distinguish sub/super-Eddington accretion, six objects belong to super-Eddington accretion category representing roughly 30\% of our sample. We find no apparent difference between these two categories in the $R_{\rm BLR}$--$R_{\rm CER}$ relation within our sample. However, due to the limited sample size and quality of lag measurement, the dependence on accretion properties need to be studied more thoroughly with future observations of a larger sample.

\section{Discussion}\label{sec:diss}

CRM approach is a new tool of BH mass estimation. Compared to the traditional methods of BH mass estimation, i.e., the spectroscopic BLR RM and SE BH mass, it has several unique advantages:

\begin{itemize}
    \item First, the broad band photometry is much more efficient compared to the time consuming spectroscopic RM. It can be easily applied to a large sample of AGNs with large-area time-domain survey, e.g., ZTF, Dark Energy Survey \citep{DEScollaboration16}, and the upcoming LSST \citep{Ivezic19} conducted by the Vera Rubin Observatory. It doesn't need further internal calibration process, i.e., using \OIII\ \citep{vanGroningen92}, which makes CRM BH mass easy to obtain.

    \item This approach provides a promising opportunity to measure BH mass for high redshift AGNs since CRM requires a much shorter monitoring baseline. For example, the $g$-$i$ lags of the 9 AGNs (log$L_{5100}$$\sim$44.5 erg s$^{-1}$, $z$$\sim$0.1) from core and parent sample range from 3 to 13 days in the observed-frame, roughly 1/5 to 1/14 of \Hbeta\ lags. An AGN at $z=2$ with similar luminosity will have an expected observed-frame continuum lag of less than 40 days. Thus, a 120-day baseline will be sufficient to recover its continuum lag, much shorter than the requirement of broad emission-line RM \citep[e.g.,][]{Grier19}. It is also possible to measure the CRM BH mass of an AGN at $z = 6$ using near-infrared photometric monitoring if the variability is not fully smoothed by time dilation. As estimated, the inter-band lag between $J$ and $K$ is $\sim80$ days in the observed frame for a quasar with log$L_{\rm bol}$$\sim$47.0 erg s$^{-1}$, which can be recovered with a $\gtrsim180$ days monitoring using $\gtrsim6$m telescope.

    \item It also allows small telescopes (aperture size $1\sim2$~m) to measure the mass of small BHs. For example, NGC 4395 contains an intermediate-mass BH, and exhibits low continuum luminosity and very weak broad emission lines. It needs 8m-level telescopes for spectroscopic RM campaign \citep{Cho21}, while its CRM BH mass is still accessible using a 2m telescope as in the case of the successful measurement of minute-level continuum lags by \cite{Montano22}. Such analysis can be extended to a large variability selected IMBH sample \citep[e.g.,][]{Baldassare20,Ward22,Burke22, Shin22,Treiber22} with well-sampled light curves. In our work, we find that the ratio $R_{\rm BLR}/R_{\rm CER}$ of NGC~4395 seems consistent with higher luminosity AGNs, which further paves the way to extend the CRM method to the IMBH regime. 

    \item The BH mass obtained from the $R_{\rm BLR}$--$R_{\rm CER}$ relation can be potentially less biased than the SE BH mass based on the $R_{\rm BLR}$--$L$ relation which is affected by the accretion rate and the host light. On the other hand, the $R_{\rm CER}$ is a direct size measurement and both the diffuse continuum and broad emission-line originate from the BLR sharing similar kinematics. The tight $R_{\rm BLR}$-$R_{\rm CER}$ relation presented in this work demonstrates the feasibility of direct BH mass estimation through CRM despite using a SE line width. Moreover, the intrinsic scatter of the relation can be further reduced with simultaneous $R_{\rm BLR}$ and $R_{\rm CER}$ measurements. These expectations can be tested with a larger sample.
    
    \end{itemize}

On the other hand, CRM BH mass also has several caveats. As noticed in \S \ref{sec:results}, the predicted $R_{\rm BLR}$ of some objects in the parent sample shows large offsets. It may suggest that the current quality of the light curves, i.e., cadence and number of bands, are not enough to accurately determine the $R_{\rm CER, BLR}$. In addition, the DC contribution to the continuum lag depends on the continuum luminosity (G22). This correlation probably caused by the Baldwin effect of the DC \citep{Li21} will also systematically affect the observed $R_{\rm CER}$--$L$ relation. The study of such secondary dependence on different AGN properties is beyond the scope of this work. Apart from these, it is known that some objects do not show clear u/U band excess \citep{Kara21,McHardy23}, indicating a small diffuse continuum contribution to the continuum lags. At this point, its influence on our $R_{\rm CER}$--$L$ relation is still unclear as separating the DC component from the disk is not easy \citep{Guo22c}.  

The next-generation time-domain survey, i.e., LSST, will help solve some of these problems and provide more accurate BH mass estimation. Its main survey, the Wide-Fast-Deep survey (WFD), will have six filters ($ugrizy$) covering from optical to near-infrared wavelength range. Any footprint in its $\sim$18000 square degree sky coverage will be observed $\sim 1000$ times in the baseline of 10 years. The single visit can reach a depth of 24.5 mag in $r$ band, and the expected number of AGNs monitored by WFD will be $6.2\times 10^{6}$ \citep{DeCicco21}. In addition to WFD, LSST will allocate $\sim10$ \% of its observing time to five Deep-Drilling-Field (DDF) for a high-cadence ($\sim 14000$ visits) deep monitoring survey down to a coadded depth of $ugrizy$ deeper than $26.5\sim28.5$ mag \citep{Brandt18}. The accuracy of the lags can be 5\% or 15\% with a 2 or 5-day cadence, respectively \citep{Nunez23}. As studied in \citet{Kovacevic22}, if the observational strategy for these fields has a cadence of $1\sim5$ days and duration $>9$ years, the expected number of sources, whose continuum lags can be successfully measured, is $>$1000 for each DDF (10 square degree) in any filter. Therefore, the DDFs will become ideal laboratories to measure CRM BH mass as well as investigate the possibility of disentangling the $R_{\rm BLR}$ \citep{Czerny23} and $R_{\rm CER}$ through the photometric light curves.

Finally, it needs to mention that the CRM approach needs a SE spectrum to provide the velocity information. Using non-simultaneous observation of spectrum and CRM will provide additional uncertainties of the BH mass. Therefore, large-area spectroscopic surveys that will accompany the LSST, e.g., the 4-meter Multi-Object Spectroscopic Telescope (4MOST) \citep{4MOST19} and the Manuakea Spectroscopic Explorer (MSE) \citep{MSE19}, will benefit our method and provide more accurate BH mass estimations.

\section{Conclusions}\label{sec:con}

In this Letter we present a new BH mass estimator via optical continuum RM. Based on the similarity between the BLR and the CER size--luminosity relations \citep[][G22]{Netzer22}, we find a tight scaling relation between the size of BLR and CER using a sample of 21 AGNs. We find that the BLR size is about 8.1 larger than that of CERs at 5100\AA\ (Figure \ref{fig:R-L}), consistent with the model prediction from \cite{Netzer22}. The intrinsic scatter of this relation is about 0.28 dex, comparable to the \Hbeta\ $R_{\rm BLR}-L$ relation (0.22 dex). We apply this relation to  the objects with reliable CER size measurements but without BLR size measurements to estimate the BLR size and BH mass (Figure \ref{fig:appliation-R-L}). We find that the predicted $R_{\rm BLR}$ of these objects follow the existing \Hbeta\ $R_{\rm BLR}$--$L$ relation well,  and the estimated BH mass is also consistent with the RM/SE BH mass using \Hbeta. Our proposed continuum RM BH mass approach will play an important role in the era of LSST. With the advent of more accurate and high-cadence light curves, the new continuum $R$--$L$ relation will allow us to estimate the BH mass for a large sample of AGNs, especially for the IMBHs and high-redshift AGNs.

\begin{acknowledgments}
We thank the anonymous referee for constructive comments that improved the manuscript. We thank Mouyuan Sun and Jinyi Shangguan for helpful suggestions and discussion. This work is supported by the National Key R\&D Program of China No.2022YFF0503402, the National Research Foundation of Korea (NRF) grant funded by the Korean government (MEST) (No. 2019R1A6A1A10073437), and the Basic Science Research Program through the National Research Foundation of Korean Government (2021R1A2C3008486).

\end{acknowledgments}

\vspace{5mm}

\software{AstroPy \citep{AstropyCollaboration18}, {\tt emcee}\citep{Foreman-Mackey13}}

\bibliography{ref.bib}\label{sec:ref}
\bibliographystyle{aasjournal}

\end{document}